\documentclass[aps,reprint,twocolumn,superscriptaddress,nofootinbib]{revtex4-2}
\usepackage[caption=false]{subfig}
\usepackage{amsmath,amsfonts,amsthm,graphicx,mathtools,epstopdf,array,etoolbox,makecell}
\usepackage[T1]{fontenc}

\usepackage{pdfpages}
\makeatletter 
\AtBeginDocument{\let\LS@rot\@undefined}
\makeatother
\usepackage{pgffor} 

\usepackage{hyperref} 
\usepackage{xurl} 

\usepackage{tikz}
\usetikzlibrary{decorations,snakes,decorations.markings,positioning,shadows.blur,calc,knots,shapes.misc}

\usepackage{xcolor}
\definecolor{LightGray}{RGB}{220,220,220}
\definecolor{myred}{RGB}{255, 19, 0}
\definecolor{myblue}{RGB}{14, 81, 167}
\definecolor{myorange}{RGB}{255, 129, 0}
\definecolor{mygreen}{RGB}{0, 146, 44}

\hypersetup{
breaklinks=true,
colorlinks = true,
citecolor= myblue,
urlcolor= myblue,
linkcolor = myblue
}

\newcommand{\tr}[2][]{\operatorname{Tr}_{#1}\!\left[#2\right]} 
\newcommand{\binh}{h_2} 

\newcommand{\defvar}{\coloneqq} 

\newcommand{\id}{\mathbb{I}} 

\newcommand{\suchthat}{\text{ s.t.}} 
\newcommand{\term}[1]{\textit{#1}}

\newcommand{\qA}{Q} 
\newcommand{\qBfar}{\widetilde{Q}} 
\newcommand{\qBnear}{\widehat{Q}} 
\newcommand{\qE}{E}
\newcommand{\cA}{A} 
\newcommand{\cB}{B} 
\newcommand{\cX}{X}
\newcommand{\cY}{Y}
\newcommand{\cP}{\cX \cY} 

\newcommand{\opA}{M_{\qA}} 
\newcommand{\opBfar}{\widetilde{M}_{\qBfar}}
\newcommand{\opBnear}{\widehat{N}_{\qBfar \qE}}

\newcommand{\Phiph}{\Phi_{\mathrm{qPER}}}
\newcommand{\PhiCHSH}{\widehat{\Phi}_{\mathrm{CHSH}}}
\newcommand{\phiph}{\phi_{\mathrm{qPER}}}
\newcommand{\phiCHSH}{\widehat{\phi}_{\mathrm{CHSH}}}

\theoremstyle{definition} 

\newcommand{\appGEAT}{Supplement~A~\cite{supp}}
\newcommand{\appreductions}{Supplement~B.1~\cite{supp}}
\newcommand{\appBFF}{Supplement~B.2~\cite{supp}}
\newcommand{\appnumerics}{Supplement~B.3~\cite{supp}}

\newcommand{\optional}[1]{#1}

\begin{document}

\title{Entropy bounds for device-independent quantum key distribution\\ with local Bell test}

\author{Ernest Y.-Z.\ Tan}
\email{yzetan@uwaterloo.ca}
\affiliation{Institute for Quantum Computing and Department of Physics and Astronomy, University of Waterloo, Waterloo, Ontario N2L 3G1, Canada}
\author{Ramona Wolf}
\email{rawolf@phys.ethz.ch}
\affiliation{Institute for Theoretical Physics, ETH Z\"{u}rich, 8093 Z\"{u}rich, Switzerland}
\affiliation{Naturwissenschaftlich-Technische Fakultät, Universität Siegen, 57068 Siegen, Germany}

\begin{abstract}
\fontfamily{lmr}\selectfont
One of the main challenges in device-independent quantum key distribution (DIQKD) is achieving the required Bell violation over long distances, as the channel losses result in low overall detection efficiencies. Recent works have explored the concept of certifying nonlocal correlations over extended distances through the use of a local Bell test. Here, an additional quantum device is placed in close proximity to one party, using short-distance correlations to verify nonlocal behavior at long distances. However, existing works have either not resolved the question of DIQKD security against active attackers in this setup, or used methods that do not yield tight bounds on the keyrates. In this work, we introduce a general formulation of the keyrate computation task in this setup that can be combined with recently developed methods for analyzing standard DIQKD. Using this method, we show that if the short-distance devices exhibit sufficiently high detection efficiencies, positive keyrates can be achieved in the long-distance branch with lower detection efficiencies as compared to standard DIQKD setups. This highlights the potential for improved performance of DIQKD over extended distances in scenarios where short-distance correlations are leveraged to validate quantum correlations.
\end{abstract}

\maketitle

\fontfamily{lmr}\selectfont 

\section{Introduction}

Quantum Key Distribution (QKD) \cite{Bennett1984,Eke91} is the only technology that, in principle, enables fundamentally unbreakable encryption. Nevertheless, its practical application faces susceptibility to attacks at the implementation level, which exploit deviations between the theoretical device models and their physical realizations \cite{GLL+11,LCT14,SK14,DLQY16,XMZ20,Zapatero2023}. To address this vulnerability, researchers have developed Device-Independent (DI) QKD \cite{BHK05,PAB+09,Sca12,PGT23}. With this approach, security no longer depends on the precise characterization of the quantum devices, but can be guaranteed solely by analyzing their classical inputs and outputs. However, its practical implementation hinges on successful execution of a Bell test free of security loopholes, making it particularly sensitive to noise. Consequently, state-of-the-art implementations are constrained to relatively short transmission distances and limited key generation rates \cite{NDN+22,ZLR+22,LZZ+22}. This illustrates that while promising, DIQKD still has to go a long way until it will be of practical use. 

\begin{figure}[t]
	\centering
	\begin{tikzpicture}[scale=0.925]
		\draw[dashed, thick, draw=black!70, rounded corners,fill=LightGray!25] (-2.9,-1.6) rectangle (3.9,1.6);
		\node[color=black!70] at (-1.7,1.85) {\small \textsf{local Bell test}};
		\draw[fill=LightGray] (0,0) circle(0.4cm);
		\node at (0,0) {\large\textsf{S}};
		\draw[fill=myred!30,draw=black, rounded corners] (-2.5,-0.5) rectangle (-1.5,0.5);
		\node at (-2,0) {\large\textsf{A}};
		\draw[->,>=stealth] (-0.4,0) -- (-1.45,0);
		\draw[->,>=stealth]	(-2,1) -- (-2,0.55);
		\draw[->,>=stealth]	(-2,-0.5) -- (-2,-0.95);
		\node at (-2,1.15) {$x$};
		\node at (-2,-1.15) {$a$};
		\draw (0.4,0) -- (1.1,0);
		\node at (1.4,-0.5) {\scriptsize\textsf{switch}};
		\draw[fill=white] (1.1,-0.3) rectangle (1.7,0.3);
		\draw[->,>=stealth] (1.7,0) -- (2.45,0);
		\draw[->,>=stealth] (1.1,0) -- (1.7,0);
		\draw[->,>=stealth,color=black] (1.1,0) to[bend right] (1.4,0.3);
		\draw[fill=myblue!30,draw=black, rounded corners] (2.5,-0.5) rectangle (3.5,0.5);
		\node at (3,0) {\large$\widehat{\textsf{B}}$};
		\draw[->,>=stealth]	(3,1) -- (3,0.55);
		\draw[->,>=stealth]	(3,-0.5) -- (3,-0.95);
		\node at (3,1.2) {$\hat{y}$};
		\node at (3,-1.15) {$\hat{b}$};
		\begin{knot}
			\strand [looseness=1, color=black] (1.4,0.3) to [out=up,in=down] (1.4,1.25) to [out=up, in=left] (2.15,2) to [out=right,in=left] (4,2);
		\end{knot}
		\draw[->,>=stealth,color=black] (4,2) -- (4.75,2);
		\begin{scope}[xshift=2.25cm,yshift=2cm]
			\draw[fill=myblue!30,draw=black, rounded corners] (2.5,-0.5) rectangle (3.5,0.5);
			\node at (3,0) {\large\textsf{B}};
			\draw[->,>=stealth]	(3,1) -- (3,0.55);
			\draw[->,>=stealth]	(3,-0.5) -- (3,-0.95);
			\node at (3,1.15) {$y$};
			\node at (3,-1.15) {$b$};
			\draw (1,0.4) circle(0.4cm);
			\draw (1.1,0.4) circle(0.4cm);
			\draw (1.2,0.4) circle(0.4cm);
		\end{scope}
	\end{tikzpicture}
	\caption{\label{fig:setting} Setting for DIQKD with a local Bell test. Alice (\textsf{A}) and Bob (\textsf{B}) want to establish a secret key over a long distance with the help of a local Bell test performed by Alice and Ben ($\widehat{\textsf{B}}$).}
\end{figure}

In response to this challenge, a novel approach was introduced that aims at certifying quantum correlations over extended distances \cite{LPT+13,CVP24}. In addition to the usual QKD setting, where two spatially separated parties (Alice and Bob) aim to establish a shared secret key, this innovative architecture employs a local Bell test conducted in close proximity to Alice, with another party we shall call Ben. The setting is illustrated in Fig.~\ref{fig:setting}. In every round of the protocol there is a random choice between performing the local Bell test with Ben, versus transmitting the quantum state to Bob (visualized with a switch in the figure). The advantage of this setup is that no particular distance is required between Alice and Ben, as long as a no-communication condition can still be enforced between the devices (for instance via appropriate ``shielding'', which would anyway be needed to keep the raw keys secret).
Hence, one can minimize all losses that occur due to long distances and thus achieve a large Bell violation. As a result, the local Bell test allows self-testing of the devices close to the source, including Alice's measurements. This information can be helpful in mitigating potential eavesdropper attacks on the DIQKD protocol between Alice and Bob. 

The idea behind this architecture was initially proposed in \cite{LPT+13}, though the original work employed an entanglement-swapping procedure rather than the routed Bell test configuration depicted in Fig.~\ref{fig:setting}. Subsequently, it was observed in \cite{CVP24} that the entanglement-swapping procedure could be effectively substituted with the routed configuration. In \cite{CVP24}, the authors studied the question of how to certify nonlocality in this setup, using the standard CHSH setting. They argue that for any non-zero CHSH violation in the local Bell test, 
some form of nonlocality in the distant branch can be certified even 
without a CHSH violation in that branch.
This result suggests that routed Bell test setups could have the potential to significantly extend the range over which nonlocality can be demonstrated, a crucial foundation for carrying out a DIQKD protocol. However, as pointed out in \cite{Lobo2023}, the form of nonlocality certified by~\cite{CVP24} is somewhat weak; furthermore, the analysis does not account for active attackers in the long-distance branch, which would be essential in any QKD security proof.

With the exception of \cite{LPT+13}, prior investigations of this setup have not analyzed the security of DIQKD against an active attacker. In \cite{LPT+13}, the security analysis employed an indirect approach that first characterizes Alice's measurements through the local Bell test, followed by keyrate computation using one-sided DI proof techniques. However, because of this intermediate step of characterizing Alice's measurements, this method results in somewhat sub-optimal keyrates. 

Our contribution in this work lies in introducing a direct formulation of the keyrate computation task in this setup without going via one-sided DI techniques, allowing us to apply recently developed methods~\cite{BFF24} for keyrate computations in standard DI setups. With this approach, we obtain a flexible method to compute much tighter bounds on the keyrates in such DIQKD configurations, avoiding the sub-optimality of the approach in~\cite{LPT+13}.
Applying our method to some demonstrative scenarios, we can show that if the detection efficiencies are sufficiently high in the short-distance devices, then it is possible to obtain positive keyrates with lower detection efficiencies in the long-distance branch as compared to standard DIQKD setups. 

\section{Methods}
\label{sec:methods}

In this work we focus only on computing the asymptotic keyrates in the limit of a large number of rounds. However, the methods we present can be directly extended to prove finite-size security against arbitrary attacks 
by using the \term{generalized entropy accumulation theorem} (GEAT)~\cite{MFSR22}. We briefly describe this in~{\appGEAT}, though it is not the focus of this work. 

We focus on protocols with the structure described in~\cite{MFSR22}, as follows. The protocol consists of $n$ rounds where states are generated and sent to the parties, who supply classical inputs to their devices and obtain classical outputs. For each round, it is randomly chosen 
to be either a \term{generation round} or a \term{test round}. In the former, the routed register is always sent to Bob, and Alice and Bob's outputs are stored for key generation later. In the latter, the routed register is randomly sent (under some distribution) to either Bob or Ben, and the outputs are later announced. 
The parties then check whether the outcome frequencies in these announced values are ``sufficiently close'' (see~\cite{MFSR22}) to the expected behavior, to decide whether to abort. 
If they do not abort, Alice and Bob then perform some classical postprocessing steps (\optional{error correction and privacy amplification;} see~\cite{rennerthesis,MFSR22}) to produce their final keys.

Focusing on a single round, we can describe the processes involved as follows. 
First, a state $\sigma_{\qA\qBnear\qE}$ is prepared, where $\qE$ is held by Eve, $\qA$ is measured in Alice's device, and $\qBnear$ is sent to either Ben or Bob. If it is sent to Ben, his device simply measures $\qBnear$. If it is sent to Bob, Eve applies some \emph{arbitrary} attack channel $\qBnear\qE \to \qBfar\qE$ (which we can take to be an invertible isometry~$V$ without loss of generality, via a suitable Stinespring dilation), then forwards $\qBfar$ to be measured in Bob's device---we denote this state produced by $V$ as $\rho_{\qA\qBfar\qE} \defvar (\id_{\qA} \otimes V)\sigma_{\qA\qBnear\qE} (\id_{\qA} \otimes V^\dagger)$.
Denote the POVM elements describing Alice's measurement on input $x$ as $\{\opA^{a|x}\}_a$; analogously denote POVM elements for Bob and Ben as $\{\opBfar^{b|y}\}_b$ and $\{\widehat{M}_{\qBnear}^{\hat{b}|\hat{y}}\}_{\hat{b}}$ respectively.

To obtain the keyrate according to the GEAT~\cite{MFSR22}, the main task to address is solving the following optimization for $\phi_{abxy}, \widehat{\phi}_{a\hat{b}x\hat{y}} \in \mathbb{R}$, where the constraints are indexed by all $a,b,\hat{b},x,y,\hat{y}$ values that occur in test rounds:
\begin{align}\label{eq:baseopt}
\begin{gathered}
\inf_{\sigma_{\qA\qBnear\qE},V,\opA^{a|x},\opBfar^{b|y},\widehat{M}_{\qBnear}^{\hat{b}|\hat{y}}}  H(\cA|\cP\qE)_{\mathrm{gen}}\\
\begin{aligned}
\suchthat \;
&\tr{\rho_{\qA\qBfar} \left(\opA^{a|x} \otimes \opBfar^{b|y}\right)} = \phi_{abxy}, \\ 
&\tr{\sigma_{\qA\qBnear} \left(\opA^{a|x} \otimes \widehat{M}_{\qBnear}^{\hat{b}|\hat{y}}\right)} = \widehat{\phi}_{a\hat{b}x\hat{y}}, \\
\text{where }& \rho_{\qA\qBfar\qE} \defvar (\id_{\qA} \otimes V)\sigma_{\qA\qBnear\qE} (\id_{\qA} \otimes V^\dagger),
\end{aligned}
\end{gathered}
\end{align}
with $H(\cA|\cP\qE)_{\mathrm{gen}}$ denoting the conditional von Neumann entropy~\cite{NC10} of Alice's outputs against Eve in a generation round, 
where $\cP$ are registers recording Alice and Bob's inputs in such rounds. Asymptotically, we can interpret $\phi_{abxy}, \widehat{\phi}_{a\hat{b}x\hat{y}}$ as the values produced by the honest behaviour---the qualitative idea is that in the IID asymptotic limit, the protocol only accepts strategies that reproduce the honest 
correlations up to a vanishingly small tolerance, hence it suffices to bound the worst-case $H(\cA|\cP\qE)_{\mathrm{gen}}$ value over such strategies. (See~\cite{MFSR22} or~{\appGEAT} for elaboration and non-IID finite-size analysis.)

The challenge in analyzing this optimization is that $V$ is only applied when the state is sent to Bob.
We resolve this challenge by reparametrizing the optimization in a \emph{completely equivalent} way, using $\rho_{\qA\qBfar\qE}$ to ``merge'' $\sigma_{\qA\qBnear\qE}$ and $V$. First, note that the objective function is already a function of $\rho_{\qA\qBfar\qE}$, since this is the state produced by Eve's attack $V$ in generation rounds.
Also, if we define some POVM elements $\opBnear^{\hat{b}|\hat{y}} \defvar V \left(\widehat{M}_{\qBnear}^{\hat{b}|\hat{y}} \otimes \id_{\qE}\right) V^\dagger$, then we have  $\tr{\sigma_{\qA\qBnear} \left(\opA^{a|x} \otimes \widehat{M}_{\qBnear}^{\hat{b}|\hat{y}}\right)} = \tr{\rho_{\qA\qBfar\qE} \left(\opA^{a|x} \otimes \opBnear^{\hat{b}|\hat{y}}\right)}$. With this we see the optimization~\eqref{eq:baseopt} has an equivalent reformulation:
\begin{align}
\label{eq:mainopt}
\begin{gathered}
\inf_{\rho_{\qA\qBfar\qE},\opA^{a|x},\opBfar^{b|y},\opBnear^{\hat{b}|\hat{y}}} \; H(\cA|\cP\qE)_{\mathrm{gen}}\\
\begin{aligned}
\suchthat \;
&\tr{\rho_{\qA\qBfar} \left(\opA^{a|x} \otimes \opBfar^{b|y}\right)} = \phi_{abxy}, \\ 
&\tr{\rho_{\qA\qBfar\qE} \left(\opA^{a|x} \otimes \opBnear^{\hat{b}|\hat{y}}\right)} = \widehat{\phi}_{a\hat{b}x\hat{y}}, 
\end{aligned}
\end{gathered}
\end{align}
with $\rho_{\qA\qBfar\qE}$ treated as an arbitrary state, and $\opBnear^{\hat{b}|\hat{y}}$ as arbitrary POVMs. (This means we do not retain the $V \left(\widehat{M}_{\qBnear}^{\hat{b}|\hat{y}} \otimes \id_{\qE}\right) V^\dagger$ structure---this clearly yields valid lower bounds by giving ``more power'' to the dishonest behaviour, but in~{\appreductions} we furthermore show that there is no loss of tightness here, and also discuss some other subtleties.)

To solve the optimization~\eqref{eq:mainopt}, we apply methods developed in~\cite{BFF24}.  
Specifically, it was shown in that work that such optimizations can be 
lower-bounded using a sequence of semidefinite programs (SDPs) known as the \term{NPA hierarchy}~\cite{NPA08} (details in~{\appBFF}). The method can also accommodate the technique of~\term{noisy preprocessing}~\cite{RGK05,KGR05,RS07,HST+20,WAP21,SBV+21}, in which trusted noise is added to the raw key to potentially improve the keyrates. 
Therefore, the perspective we have introduced allows us to compute the keyrates for a wide variety of protocols, by implementing this method to tackle the optimization~\eqref{eq:mainopt}.

Given a lower bound $h\in\mathbb{R}$ on the optimization~\eqref{eq:mainopt} for the case where $\widehat{\phi}_{a\hat{b}x\hat{y}}$ are the values produced by honest behaviour, the asymptotic keyrate is given by the ``Devetak-Winter formula''~\cite{DW05} (originally derived in the IID case; see e.g.~\cite{MFSR22} for a proof in the non-IID case):
\begin{align}
\label{eq:keyrate}
h - H(\cA|\cP \cB)^\mathrm{hon}_\mathrm{gen},
\end{align}
where $H(\cA|\cP \cB)^\mathrm{hon}_\mathrm{gen}$ is the conditional entropy between Alice and Bob's generation-round outputs (after noisy preprocessing, if applied) in the honest behaviour.
\optional{Qualitatively, the $h$ term quantifies Eve's uncertainty about Alice's raw outputs, while the $H(\cA|\cP \cB)^\mathrm{hon}_\mathrm{gen}$ term quantifies how much Eve later learns during error correction.} We use this formula in our subsequent keyrate calculations.

\section{Results}

We apply our method to a routed Bell setup with inputs $x\in\{0,1\}$, $y\in\{0,1,2,3\}$, $\hat{y}\in\{0,1\}$, where for generation rounds the fixed input pair $(x,y)=(0,3)$ is always used, whereas for test rounds all the inputs except $y=3$ are used. This means that when we evaluate the optimization~\eqref{eq:mainopt}, we do not include constraints with $y=3$; this reduces the SDP size (see~{\appnumerics}) while still providing positive keyrates over a large range of detection efficiencies. 

The idea behind this choice is that Ben can use his two inputs to certify short-distance CHSH violation, whereas for Bob, loosely speaking he could use input values $y\in\{0,1\}$ to certify long-distance CHSH violation, and also use $y=2$ to certify a ``quasi-phase-error rate''\footnote{It is not a phase error rate in the standard sense of~\cite{Koa09,MPW22}, because Alice's true $x=1$ measurement may not be conjugate to $x=0$.} (qPER) with Alice's $x=1$ input. Recall that given our earlier discussions, if the short-distance CHSH violation is maximal, then in principle Bob should only need to certify the qPER and not the CHSH violation. However, we find (see Fig.~\ref{fig:fullprob}) that once we consider imperfect detectors, it is significantly better for Bob to have enough input values to certify \emph{both} his qPER and CHSH value with respect to Alice---in particular, this ensures the resulting keyrates are always at least as good as the ``usual'' CHSH-inspired DIQKD setups. 

We consider the following noise model. We suppose that Alice/Ben/Bob have limited detection efficiencies $\eta_A,\eta_B,\eta_{\widehat{B}}\in[0,1]$, respectively. With this, we model the honest devices as being able to generate any specified two-qubit state 
and perform any specified projective measurement with outcomes $\{0,1\}$; however, for each device (independently) the outcome is replaced with a ``no-detection'' value $\bot$ with probability $1-\eta_P$ where $\eta_P$ is that party's detection efficiency. 
\optional{We stress that this model is only used to determine the values of $\phi_{abxy}$ and $\widehat{\phi}_{a\hat{b}x\hat{y}}$ attainable in the honest case---when bounding the optimization~\eqref{eq:mainopt}, the infimum is evaluated over all possible adversarial states and measurements, with no dimension restriction.}
To make the SDP sizes tractable, we suppose the outcomes are then ``coarse-grained'' by mapping the $\bot$ outcome to the $0$ outcome, \emph{except} when Bob uses input $y=3$, in which case we follow~\cite{ML12} and preserve the $\bot$ value to improve the keyrates.
(As noted above, we do not include the $y=3$ terms in the constraints, hence this does not affect the SDP size; see~{\appnumerics}.)
In summary, we always have $a,b,\hat{b}\in\{0,1\}$, except when $y=3$, in which case $b\in\{0,1,\bot\}$ instead.

As a starting demonstration, we solve a relaxed version of the optimization~\eqref{eq:mainopt} with ``coarse-grained'' constraints; specifically:
\begin{align}
\label{eq:relaxedopt}
\begin{gathered}
\inf_{\rho_{\qA\qBfar\qE},\opA^{a|x},\opBfar^{b|y},\opBnear^{\hat{b}|\hat{y}}} \; H(\cA|\cP\qE)_{\mathrm{gen}}\\
\begin{aligned}
\suchthat \;
&\tr{\rho_{\qA\qBfar} \Phiph\!\left(\opA^{a|x},\opBfar^{b|y}\right)} = \phiph, \\ 
&\tr{\rho_{\qA\qBfar\qE} \PhiCHSH\!\left(\opA^{a|x},\opBnear^{\hat{b}|\hat{y}}\right)} = \phiCHSH, 
\end{aligned}
\end{gathered}
\end{align}
where
\begin{align} \label{eq:2paramsops}
\begin{gathered}
\Phiph\!\left(\opA^{a|x},\opBfar^{b|y}\right) \defvar \sum_{\substack{a,b \in \{0,1\} \\ \suchthat\; a\neq b}} \opA^{a|1} \otimes \opBfar^{b|2}, \\
\PhiCHSH\!\left(\opA^{a|x},\opBnear^{\hat{b}|\hat{y}}\right) \defvar 
\sum_{\substack{a,\hat{b},x,\hat{y} \in \{0,1\} \\ \suchthat\; a\oplus\hat{b} = x\hat{y}}} \frac{1}{4} \opA^{a|x} \otimes \opBnear^{\hat{b}|\hat{y}}
,
\end{gathered}
\end{align}
and the constraint values $\phiph,\phiCHSH \in \mathbb{R}$ are the expected asymptotic values from the honest behavior. 
Qualitatively, $\phiph$ describes the qPER between Alice and Bob, 
while $\phiCHSH$ describes the CHSH game winning probability between Alice and Ben.
In this starting example, we are ignoring Bob's inputs $y \in \{0,1\}$: this lets us verify the intuition that if 
$\phiCHSH$ is large enough, then 
$\phiph$ should suffice to certify nontrivial entropy bounds, without a distant Bell violation. 

\begin{figure}[t!]
\centering
\includegraphics[width=0.49\textwidth]{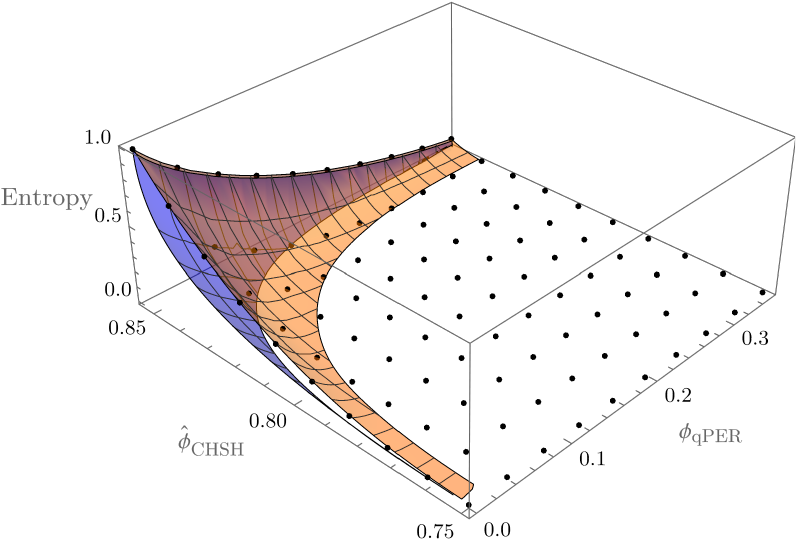}
\caption{Our bounds on the optimization~\eqref{eq:relaxedopt}, where the only constraints are the distant qPER $\phiph$ and the local CHSH value $\phiCHSH$ (see~\eqref{eq:2paramsops}). The dots show our raw data points, while the orange surface is a fitted function for those data points (where to err on the side of reliability we have only plotted it on a restricted domain where its value is at least $10^{-4}$). For comparison, the blue surface is the closed-form bound~\eqref{eq:2paramsbnd} from~\cite{TH13,LPT+13}. Our results are clearly always an improvement over that bound (with equality in the regime where it is tight, namely whenever $\phiCHSH$ is maximal); also, we obtain positive values on a significantly larger region. }
\label{fig:2params}
\end{figure}

In Fig.~\ref{fig:2params}, we plot the results of the optimization~\eqref{eq:relaxedopt} as a function of $\phiph$ and $\phiCHSH$ (without noisy preprocessing). For this scenario,~\cite{TH13,LPT+13} derived the following closed-form bound on~\eqref{eq:relaxedopt}: 
\begin{align}\label{eq:2paramsbnd}
-\log_2\left(\frac{1}{2}+\frac{\beta}{8}\sqrt{8-\beta^2}\right) - \binh(\phiph),
\end{align}
where $\beta \defvar 8\phiCHSH-4$ is the ``correlator CHSH value'' and $\binh$ is the binary entropy function.
We display this bound in Fig.~\ref{fig:2params} as well, from which it can be seen that our approach yields better results. 
Furthermore, we highlight that when $\phiCHSH$ is at its maximal value $\phiCHSH^\mathrm{max} \defvar (2+\sqrt{2})/4 \approx 
0.853
$,
the bound~\eqref{eq:2paramsbnd} reduces to $1-\binh(\phiph)$, as expected from self-testing. This particular case of the bound is tight~\cite{TH13,LPT+13}, and our approach reproduces this result, as seen from the $\phiCHSH=\phiCHSH^\mathrm{max}$ boundary in Fig.~\ref{fig:2params} (details in~{\appnumerics}).

\begin{figure*}
\subfloat[]{\label{subfig:same_eta}
\includegraphics[width=0.49\textwidth]{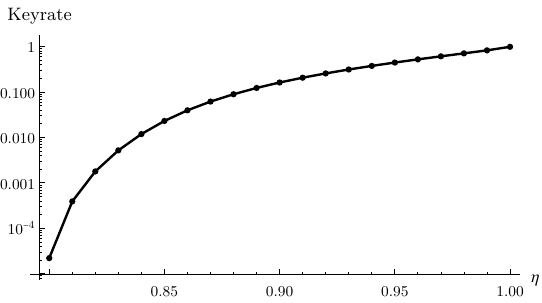}
} 
\subfloat[]{\label{subfig:diff_etaB}
\includegraphics[width=0.49\textwidth]{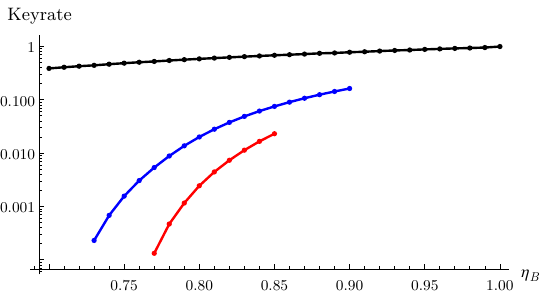}
}
\caption{In~\hyperref[subfig:same_eta]{(a)}, we plot the keyrates for a scenario where all parties have the same detection efficiency, $\eta_A=\eta_B=\eta_{\widehat{B}} \eqqcolon \eta$. In~\hyperref[subfig:diff_etaB]{(b)}, we consider a scenario where the short-distance efficiencies have some fixed values $\eta_A=\eta_{\widehat{B}} \eqqcolon \eta_\mathrm{short}$, and plot the keyrates in terms of the long-distance efficiency $\eta_B$ for $\eta_B \leq \eta_\mathrm{short}$, with the red, blue and black curves corresponding to $\eta_\mathrm{short}=0.85$, $0.90$ and $0.99999$ respectively. 
For each data point, we optimized the choice of honest states and measurements, as this can improve the keyrates~\cite{Ebe93,HST+20,WAP21,SBV+21,BFF24}; we also implement noisy preprocessing~\cite{RGK05,KGR05,RS07,HST+20,WAP21,SBV+21} and optimize the noisy preprocessing parameter value.
We see that for the $\eta_\mathrm{short}=0.85$ and $\eta_\mathrm{short}=0.90$ cases, keyrates of at least $10^{-4}$ are achievable for $\eta_B \gtrsim  0.77$ and $\eta_B \gtrsim  0.73$ respectively. 
}
\label{fig:fullprob}
\end{figure*}

Turning to detection inefficiency, we performed some heuristic exploration by treating the fitted function in Fig.~\ref{fig:2params} as a valid bound on~\eqref{eq:relaxedopt}, and finding the minimum detection efficiencies that would be needed to produce positive keyrates in~\eqref{eq:keyrate}. (For every tuple of values $(\eta_A,\eta_B,\eta_{\widehat{B}})$, we heuristically optimized the honest states and measurements to maximize the keyrate, as discussed in~\cite{Ebe93,HST+20,WAP21,SBV+21,BFF24}.
We do not consider noisy preprocessing here, because the data points in Fig.~\ref{fig:2params} do not incorporate that.) 
If Alice and Ben have perfect detectors, $\eta_A=\eta_{\widehat{B}}=1$, then the keyrates are very robust as Bob's detection efficiency decreases, 
remaining positive for $\eta_B \gtrsim 0.68$ at least. 
Unfortunately, this finding is not robust when $\eta_A,\eta_B$ are decreased: for instance, if we set all detection efficiencies equal ($\eta_A=\eta_B=\eta_{\widehat{B}} \eqqcolon \eta$), then we only obtained positive keyrates for $\eta \gtrsim 0.96$, which is worse than standard CHSH-based protocols~\cite{PAB+09}. The cause appears similar to~\cite{SGP+21}: for $\eta < 1$, the state that maximizes CHSH violation is not maximally entangled~\cite{Ebe93}; due to this, it is difficult for Alice's $x=0$ and $x=1$ measurements to \emph{both} be highly correlated to measurements on Bob's state, which creates a tension between minimizing $H(\cA|\cP \cB)^\mathrm{hon}_\mathrm{gen}$ in~\eqref{eq:keyrate} versus minimizing $\phiph$. 

However, an advantage of our approach now becomes apparent: since it can handle optimizations of the form~\eqref{eq:mainopt}, we do not have to restrict ourselves to the coarse-grained parameters $\phiph,\phiCHSH$ in~\eqref{eq:relaxedopt}, 
but can directly tackle the original optimization~\eqref{eq:mainopt} where \emph{all} the terms $\phi_{abxy}, \widehat{\phi}_{a\hat{b}x\hat{y}}$ 
are used as constraints.
Our results are shown in Fig.~\ref{fig:fullprob}. First, in Fig.~\ref{subfig:same_eta} we again consider the case where all detection efficiencies are equal, and find that now the keyrates remain positive down to $\eta=0.80$ at least (similar to~\cite{BFF24}). More importantly, the fact that they remain positive over such a range means that we can consider a scenario more relevant to implementations. Specifically, we suppose that Alice and Ben have fixed high detection efficiencies $\eta_A=\eta_{\widehat{B}} \eqqcolon \eta_\mathrm{short}$, while Bob has some lower detection efficiency $\eta_B$ (these values are to incorporate all loss mechanisms in the respective arms). For this scenario, shown in Fig.~\ref{subfig:diff_etaB}, we find that for $\eta_\mathrm{short}=0.99999$ 
the keyrates remain robust as $\eta_B$ decreases. For lower values of $\eta_\mathrm{short}$, we see that 
positive keyrates 
can be achieved for $\eta_B \gtrsim  0.77$ when $\eta_\mathrm{short}=0.85$, or $\eta_B \gtrsim  0.73$ when $\eta_\mathrm{short}=0.90$.
If we model Bob's detection efficiency in such a scenario as $\eta_B = \eta_\mathrm{short} \eta_\mathrm{chann}$ where $\eta_\mathrm{chann}$ describes the transmission efficiency in the long-distance channel, these values correspond to thresholds of $\eta_\mathrm{chann} \gtrsim 0.90$ and $\eta_\mathrm{chann} \gtrsim 0.81$ respectively. 
(See~{\appnumerics} for other variants.)

\section{Conclusion}

We introduced a numerical method to compute reliable entropy bounds for DI protocols employing a local Bell test.
Our approach yields significantly better bounds than previous results in~\cite{LPT+13}, and has the further advantage of applying to a wide range of scenarios, and incorporating techniques such as noisy preprocessing.
Applying this method to DIQKD revealed that using the local Bell test enables the certification of non-zero keyrates over longer distances (cf.~Fig.~\ref{subfig:diff_etaB}), provided that the efficiency of the local Bell test is sufficiently high. 
This represents an important step in the quest to enhance the practicality of DI protocols.

Our method could also be promising in studying local Bell tests for other protocols reliant on nonlocality as a resource. These include other (semi-)DIQKD protocols, such as the symmetric generalization suggested in \cite{CVP24}, 
or potentially some blind quantum computing protocols \cite{Reichardt2013,Gheorghiu2015,Hajdusek2015,MFSR22} if such protocols only require entropic bounds rather than self-testing properties. Therefore, it carries the potential to enhance the performance of applications leveraging Bell tests.

Similar results were independently derived in a separate work~\cite{arx_RLPP24}. 
As a brief comparison, they considered somewhat higher $\eta_\mathrm{short}$ values than this work, and also considered depolarizing noise; hence we obtain somewhat different plots. The exception is the $\eta_\mathrm{short}=1$ case, in which they explicitly found positive keyrates down to $\eta_B = 0.5$, which is a wider $\eta_B$ range than we considered in Fig.~\ref{fig:fullprob}. 
Still, the data points we showed for that case do indeed match their results in the same $\eta_B$ range, so we expect that if we extended Fig.~\ref{fig:fullprob} down to $\eta_B = 0.5$ then we would also obtain the same results. 
Another recent independent work~\cite{arx_MS24} has studied a one-sided DI scenario which assumes Alice's measurements and detectors are trusted---again, our approach should in principle reproduce their results (for perfect detection efficiency on Alice's side) by setting $\phiCHSH = \phiCHSH^\mathrm{max}$.

\begin{acknowledgments}
We thank the authors of~\cite{arx_RLPP24} for coordinating a joint release of our results.
We furthermore thank Jean-Daniel Bancal, Renato Renner, Martin Sandfuchs, and Nicolas Sangouard for helpful discussions. 
RW is supported by the Ministry of Culture and Science of North~Rhine-Westphalia (Ministerium für Kultur und Wissenschaft des Landes Nordrhein-Westfalen) via the NRW-Rückkehrprogramm. She furthermore acknowledges support from the Air Force Office of Scientific Research (AFOSR), grant No.\ FA9550-19-1-0202, the Quant\-ERA project eDICT, the National Centre of Competence in Research SwissMAP, and the Quantum Center at ETH Zurich.
EYZT conducted research for this work at the Institute for Quantum Computing, at the University of Waterloo, which is supported by Innovation, Science, and Economic Development Canada; support was also provided by NSERC under the Discovery Grants Program, Grant No. 341495.
\end{acknowledgments}

\bibliography{biblio_localBell} 



\section*{Supplementary material}

**Appendix A: Outline for finite-size security against coherent attacks**

For a detailed exposition of such security proofs, refer to e.g. [21] or the earlier works [41, 42]; here we just provide a broad outline of the key steps.

In a protocol with a finite number of rounds (denoted as $n$), we would define the accept condition to be that the observed frequency distribution in the test rounds lies within some appropriate "$\delta$-neighbourhood" of the honest expected values, where $\delta$ can be chosen such that $\delta \to 0$ as $n \to \infty$ (see e.g. [42]). We first note that this is enough to ensure that the honest devices (assumed IID) will be accepted with high probability, as we would expect.

As for a security proof, let us first gain some intuition assuming that Eve's attack is IID, following the approach in e.g. [22, 42]. We divide the space of possible IID attacks into those that (in every individual round) produce values of $\text{Tr}\left[\rho_{Q\widetilde{Q}}\left(M_Q^{a|x} \otimes \widetilde{M}_{\widetilde{Q}}^{b|y}\right)\right]$ and $\text{Tr}\left[\rho_{Q\widetilde{Q}E}\left(M_Q^{a|x} \otimes \widehat{N}_{\widetilde{Q}E}^{\hat{b}|\hat{y}}\right)\right]$ within some $\delta'$-neighbourhood of the accepted set (where again $\delta'$ can be chosen such that $\delta' \to 0$ as $n \to \infty$), and those that do not. By standard concentration inequalities, all attacks in the latter category will cause an abort with high probability, which suffices to satisfy standard security definitions [41, 42]. Hence the main focus is on attacks in the former category. We observe that the minimal value of $H(A|XYE)_\text{gen}$ over all such attacks would be simply an optimization of the form (1), except with the constraints slightly relaxed to the condition that $\text{Tr}\left[\rho_{Q\widetilde{Q}}\left(M_Q^{a|x} \otimes \widetilde{M}_{\widetilde{Q}}^{b|y}\right)\right]$ and $\text{Tr}\left[\rho_{Q\widetilde{Q}E}\left(M_Q^{a|x} \otimes \widehat{N}_{\widetilde{Q}E}^{\hat{b}|\hat{y}}\right)\right]$ lie within some small neighbourhood of the honest expected values. Given a lower bound on this optimization, one can essentially compute the finite-size keyrate by subtracting the error-correction term $H(A|XYB)_\text{gen}^\text{hon}$ as well as some explicit finite-size terms (which scale as $O(1/\sqrt{n})$) given by the quantum *asymptotic equipartition property* [43] and *leftover hashing lemma* [22]. (Some slight adjustments are needed to account for the fraction of rounds that are test rounds; we do not go further into details here.) With this, we can see that in the asymptotic $n \to \infty$ limit (where $\delta, \delta' \to 0$), the keyrate would converge to the value of the optimization (1) with $\phi_{abxy}, \widehat{\phi}_{a\hat{b}x\hat{y}}$ being exactly the honest expected values, minus the error-correction term $H(A|XYB)_\text{gen}^\text{hon}$.

To prove security against coherent attacks, we would apply the GEAT as described in [21], via the following proof sketch. First, we define a sequence of channels describing the protocol rounds, where each channel performs the processes described in the main text (e.g. state preparation, test/generation decision and routing, device measurements) and then produces a classical register $C$ storing all the input and output values from that round. Such a channel can be shown to satisfy the no-signalling condition of the GEAT between the registers $AC$ and the various "side-information" registers (as long as we include any memory in the source device as part of those side-information registers). Now let the optimal value of the optimization (1) be denoted abstractly as a function $r(\phi_{abxy}, \widehat{\phi}_{a\hat{b}x\hat{y}})$ of the constraint values $\phi_{abxy}, \widehat{\phi}_{a\hat{b}x\hat{y}}$. Using SDP methods similar to those in this work, one can derive an affine lower bound on $r(\phi_{abxy}, \widehat{\phi}_{a\hat{b}x\hat{y}})$ (for instance by using the Lagrange dual as described in [42] or [19]), and rescale the domain (see e.g. [42, 44]) such that it is a function of probability distributions on $C$ as required by the GEAT. With this, the finite-size keyrate (secure against coherent attacks) according to the GEAT is basically given by the minimal value of this affine bound over the $\delta$-neighbourhood specified in the accept condition, minus the error-correction term $H(A|XYB)_\text{gen}^\text{hon}$ and some explicit finite-size terms that scale as $O(1/\sqrt{n})$ asymptotically [21, 41, 42]. (Again, for this sketch we omit some technical details regarding the test-round fraction.) This gives similar results as in the IID case described above, though with different (typically larger) finite-size corrections.

**Appendix B: Details on optimization problem**

**1. Reductions on Ben's operators**

There is a subtlety to be aware of in our framework: when reformulating the device behaviours in terms of the new POVMs $\widehat{N}_{\widetilde{Q}E}^{\hat{b}|\hat{y}}$, we have focused on correctly reproducing the outcome probabilities in test rounds and the final states (including Eve's side-information) in generation rounds, but some inspection shows that our construction does not correctly reproduce the final states (on Eve's side-information) in test rounds where the state is routed to Ben. For the purposes of the optimization (1), this is not a problem since Eve's side-information for the test-round case does not appear anywhere in the optimization problem; furthermore, it is indeed true that solving this optimization is already sufficient to compute finite-size keyrates against either IID attacks or coherent attacks, and the asymptotic keyrates converge to the formula (2). However, when proving security against coherent attacks in DIQKD, the finite-size keyrates can be slightly improved [41, 42] by also incorporating the entropy contributions from test rounds. Our approach does not seem to give a straightforward method to compute such contributions (because it does not correctly

reproduce Eve's side-information when the state is routed to Ben), and hence it may not be easy to incorporate this slight improvement when using our approach.

We also note that in our approach, we first constructed the new POVMs from the original ones via the relation $\widehat{N}_{\widetilde{Q}E}^{\hat{b}|\hat{y}} \coloneqq V \left( \widehat{M}_{\widehat{Q}}^{\hat{b}|\hat{y}} \otimes \mathbb{I}_E \right) V^\dagger$, but in our final optimization we simply treated them as completely arbitrary POVMs on $\widehat{Q}E$. *A priori*, it would seem that this could potentially cause some looseness in the resulting bounds, because we have not preserved the constraint that they originally arose from measurements with a tensor-product structure across $\widehat{Q}$ and $E$. However, we now argue that for the purposes of bounding $H(A|XYE)_{\text{gen}}$ subject to constraints on the test-round correlations, there is in fact no loss of tightness by instead considering the optimization as expressed in (1). This is because while in the main text we allowed Eve to hold some side-information $E$ in the initial state $\sigma_{Q\widehat{Q}E}$ for the sake of generality, we can in fact show that this was not necessary, and we could have just considered an initial state of the form $\sigma_{Q\widehat{Q}}$ (with Eve's attack isometry being of the form $V: \widehat{Q} \to \widetilde{Q}E$ instead).

To do so, let us first consider some initial state $\sigma_{Q\widehat{Q}E}$ and attack isometry $V: \widehat{Q}E \to \widetilde{Q}E$ as described in the main text. Now let $\widehat{Q}'$ be a register isomorphic to $\widehat{Q}E$, and define the following device behaviour and attack by Eve: first, it generates the state $\sigma_{Q\widehat{Q}'}$ (recalling $\widehat{Q}' \equiv \widehat{Q}E$) without any register being held by Eve, then $\widehat{Q}'$ is routed to either Ben or Bob. If it is routed to Ben, his device measures $\widehat{Q}'$; if it is routed to Bob, Eve performs her attack isometry $V: \widehat{Q}' \to \widetilde{Q}E$ (recalling $\widehat{Q}' \equiv \widehat{Q}E$), and then Bob's device measures $\widetilde{Q}$. Clearly, this can produce exactly the same generation-round states and test-round outcome probabilities as the original strategy (where Eve started with some side-information in the initial state); furthermore, note that every behaviour of this form (including *any* possible initial state on $Q\widehat{Q}'$ and measurements on $\widehat{Q}'$) is clearly allowed within the routed Bell test setup we consider. However, notice that in this formulation, Eve started with no side-information in the initial state $\sigma_{Q\widehat{Q}'}$. Therefore if we now repeat our argument from the main text, the new POVMs we construct would have the form $\widehat{N}_{\widetilde{Q}E}^{\hat{b}|\hat{y}} \coloneqq V\widehat{M}_{\widehat{Q}'}^{\hat{b}|\hat{y}}V^\dagger$, without any tensor-product structure to preserve in the measurements $\widehat{M}_{\widehat{Q}'}^{\hat{b}|\hat{y}}$. (Again, this construction runs into some difficulties reproducing Eve's side-information in test rounds, so we stress that this is focused on only the generation-round states and the test-round outcome probabilities.)

### 2. Relaxation to SDPs

While it is not strictly necessary for any of the following analysis, we first note that without loss of generality, we can assume that all the measurements are projective (this property allows for some simplifications in the resulting SDPs [19, 24]). For standard Bell-test scenarios, this is a "well-known" reduction, but here we take additional care due to the interaction between Bob and Ben's measurements (in such reductions, the main subtlety as compared to a standard Naimark dilation is to ensure all the possible POVMs can be *simultaneously* replaced by projective measurements). The argument is as follows: first, note that when analyzing a single round for the purposes of applying an IID security proof or the GEAT, all quantities of interest (i.e. all relevant entropies and the probability distributions produced on various classical registers) can be written entirely in terms of the classical-quantum states produced by the various possible combination of input choices and/or routing (where the classical part of the state consists of a record of the measurement outcomes). Therefore, it would suffice to show that for any such classical-quantum states produced by general POVMs acting on some initial state, we can reproduce the same classical-quantum states by applying projective measurements on some other initial state.

To do so, we note that starting with Alice's device which could perform one of several POVMs $\{M_Q^{a|x}\}_a$, we can apply a standard Naimark dilation construction (e.g. as in [23]) to obtain the following: there exists an ancillary state $|0\rangle\langle 0|_R$ such that for each $x$, if we append $|0\rangle\langle 0|_R$ to the state on $Q$ and perform a suitable projective measurement $\{P_{RQ}^{a|x}\}_a$ across $RQ$, the outcome probabilities are the same as though we had measured $\{M_Q^{a|x}\}_a$ on $Q$ directly. (The state $|0\rangle\langle 0|_R$ does not depend on $x$, which will be critical in our later argument. If we use the alternate perspective of viewing Naimark dilations as an embedding in a larger Hilbert space, this is essentially a statement that we can perform "the same embedding" regardless of $x$.) This implies that the resulting classical-quantum states after such a process, with the classical part recording the measurement outcome, are the same as for the original POVMs. We follow an analogous construction for Bob and Ben, except that we use the same ancillary state $|0\rangle\langle 0|_{\widetilde{R}}$ for both of them, i.e. the resulting projective measurements $\{\widetilde{P}_{\widetilde{R}\widetilde{Q}}^{b|y}\}_b$ for Bob and $\{\widehat{P}_{\widetilde{R}\widehat{Q}}^{\hat{b}|\hat{y}}\}_{\hat{b}}$ for Ben act on $\widetilde{R}\widetilde{Q}$ and $\widetilde{R}\widehat{Q}$ respectively.

(For Ben, we are directly analyzing the original POVMs $\{\widehat{M}_{\widehat{Q}}^{\hat{b}|\hat{y}}\}_{\hat{b}}$, not the transformed POVMs $\{\widehat{N}_{\widetilde{Q}E}^{\hat{b}|\hat{y}}\}_{\hat{b}}$; this keeps our argument here slightly more general since it does not rely on the reductions described in the main text.) With

this, we see that for any initial state $\sigma_{Q\widehat{Q}E}$ to be measured by POVMs in Alice/Bob/Ben's devices (after the routing and/or attack channel), we can reproduce exactly the same final classical-quantum states by instead starting with the state $|00\rangle\langle 00|_{R\widetilde{R}} \otimes \sigma_{Q\widehat{Q}E}$ and having Alice/Bob/Ben's devices perform the projective measurements constructed above (implicitly using the fact that Alice's measurement commutes with Bob/Ben's). Note that for this to work, the $R$ register must be sent to Alice and the $\widetilde{R}$ register routed to either Bob or Ben (and not acted on by Eve's attack channel in the former case), but this can indeed be achieved. Put another way, we can view $RQ$ in this state as being "Alice's register" and $\widetilde{R}\widehat{Q}$ as being the register that will be routed to Bob or Ben. In conclusion, we can suppose all the parties' original measurements are projective without loss of generality. Our transformation of Ben's measurements to some new measurements $\widehat{N}^{\hat{b}|\hat{y}}_{\widetilde{Q}E} := V\left(\widehat{M}^{\hat{b}|\hat{y}}_{\widehat{Q}} \otimes \mathbb{I}_E\right) V^\dagger$ also preserves the property that they are projective.

We now turn to describing the relaxation of the optimization (1) to an SDP. In [19], a converging sequence of arbitrarily tight lower bounds were constructed for that optimization. All bounds in the sequence have the following form, where the infimum is over the state $\rho_{Q\widetilde{Q}E}$, the measurement operators $M^{a|x}_Q, \widetilde{M}^{b|y}_{\widetilde{Q}}, \widehat{N}^{\hat{b}|\hat{y}}_{\widetilde{Q}E}$, and some additional operators $Z^c_E$ acting on Eve's register $E$:

$$\begin{aligned}
\inf\ &\mathrm{Tr}\left[\rho_{Q\widetilde{Q}E} f\left(M^{a|x}_Q, Z^c_E\right)\right] \\
\text{s.t.}\ &\mathrm{Tr}\left[\rho_{Q\widetilde{Q}}\left(M^{a|x}_Q \otimes \widetilde{M}^{b|y}_{\widetilde{Q}}\right)\right] = \phi_{abxy}, \\
&\mathrm{Tr}\left[\rho_{Q\widetilde{Q}E}\left(M^{a|x}_Q \otimes \widehat{N}^{\hat{b}|\hat{y}}_{\widetilde{Q}E}\right)\right] = \widehat{\phi}_{a\hat{b}x\hat{y}},
\end{aligned} \tag{B1}$$

where the function $f\left(M^{a|x}_Q, Z^c_E\right)$ is defined in Eq. (15) in [19]: basically, it is a noncommutative polynomial in the operators $M^{a|x}_Q, Z^c_E$ (allowing the polynomial to combine tensor products and operator products). In turn, optimizations of this form can be efficiently lower-bounded using the NPA hierarchy [24]: the basic idea is to relax the above optimization by replacing all tensor products with operator products, and impose commutation relations in place of the original tensor-product structure. With regards to the above optimization, that means the resulting $M^{a|x}_Q, \widetilde{M}^{b|y}_{\widetilde{Q}}, Z^c_E$ operators after this replacement satisfy the commutation relations $\left[M^{a|x}_Q, \widetilde{M}^{b|y}_{\widetilde{Q}}\right] = \left[\widetilde{M}^{b|y}_{\widetilde{Q}}, Z^c_E\right] = \left[Z^c_E, M^{a|x}_Q\right] = 0$; also, all the $\widehat{N}^{\hat{b}|\hat{y}}_{\widetilde{Q}E}$ operators commute with all the $M^{a|x}_Q$ operators, but *not* with the $\widetilde{M}^{b|y}_{\widetilde{Q}}$ or $Z^c_E$ operators.

We computed the plots shown in this work by modifying the code provided at https://github.com/peterjbrown519/DI-rates/blob/main/2222_global.py to implement the approach we describe here. Following that code, we used global level 2 of the NPA hierarchy together with some additional monomials, and used 12 nodes in the Gauss-Radau quadrature. We also retained the simplification used in that code based on Remark 2.6 of [19], where a sum and infimum are swapped in order to significantly reduce the SDP size at the cost of potentially looser keyrate bounds. We chose to retain this in order to speed up the optimization of parameter choices for each detection efficiency value — due to the size of the input/output sets in our setup, optimizing each data point already required over an hour even with this simplification. (More precisely, any single evaluation of the optimization (B1) with this simplification is fairly fast; the time consumption arises from having to evaluate it many times to optimize the choices of $\phi_{abxy}, \widehat{\phi}_{a\hat{b}x\hat{y}}$ and noisy-preprocessing parameter.)

### 3. Numerical computation details

As noted at the start of Sec. II, we focus on protocols that do not use the $y = 3$ input in test rounds, and hence when solving the optimization (1), we omit any constraints involving this input. This implies that the $y = 3$ input entirely does not feature in the optimization (1) — while it might appear that it could play a role in the objective function due to the $Y$ register in $H(A|XYE)_{\mathrm{gen}}$, recall that we have chosen to focus on protocols where the fixed input pair $(X, Y) = (0, 3)$ is used in all generation rounds, and so we basically just have $H(A|XYE)_{\mathrm{gen}} = H(A|E)_{\mathrm{gen}}$ (though with the implicit understanding that $A$ is produced by the $X = 0$ input). Therefore, our choice to preserve the $b = \perp$ outcome label for this input choice does not affect the SDP size. Previous works such as [19] have often followed this convention of excluding this $y = 3$ input from the optimization in order to keep the SDP size manageable; it is generally believed that this choice should not cause a significant decrease in the certifiable keyrates.

We also remark that for Fig. 2, when $\phi_{\mathrm{qPER}}$ or $\widehat{\phi}_{\mathrm{CHSH}}$ are too close to their extremal values, the above SDP can be numerically unstable. Therefore, when producing the plot, the data points we chose at the plot boundaries are $10^{-5}$ away from the true extremal values. However, even with this slight deviation from the extremal values, we found

for instance that our data points for $\widehat{\phi}_{\text{CHSH}} = \widehat{\phi}_{\text{CHSH}}^{\max} - 10^{-5}$ already returned entropy bounds within 0.005 of the "self-tested" value $1 - h_2(\phi_{\text{qPER}})$ that would be obtained at $\widehat{\phi}_{\text{CHSH}} = \widehat{\phi}_{\text{CHSH}}^{\max}$ exactly; furthermore, they are in fact already slightly better than the values obtained by substituting $\widehat{\phi}_{\text{CHSH}} = \widehat{\phi}_{\text{CHSH}}^{\max} - 10^{-5}$ into the previous closed-form bound from [16, 35].

Also, regarding Fig. 3, our numerical results suggested that in fact it did not seem critical for Bob to have all 3 test-round inputs — specifically, for the $\eta_{\text{short}} = 0.85$ or $0.90$ cases, we were able to obtain almost identical keyrate plots by instead considering a scenario with only 2 test-round inputs. In other words, while our analysis of the scenario based only on $\phi_{\text{qPER}}, \widehat{\phi}_{\text{CHSH}}$ indicates that it is useful for Bob to have more test-round inputs than just the "qPER measurement" alone, it seems a *single* additional input is enough to achieve most of the advantage found in Fig. 3, rather than giving him two additional inputs to separately certify long-distance CHSH violation. Presumably, in this version Bob's inputs could be informally viewed as certifying some combination of the qPER and long-distance CHSH violation.



\end{document}